\documentclass[sigplan,nonacm,10pt]{acmart}

\renewcommand\footnotetextcopyrightpermission[1]{}
\usepackage{duckuments}
\usepackage{listings}
\usepackage{xcolor}
\usepackage{float}
\usepackage{caption}
\usepackage{tikz}
\usepackage{xspace}
\usepackage{enumitem}
\usepackage{lipsum}
\usepackage{subfig}
\usepackage[dvipsnames]{xcolor}
\usepackage{soul}
\usepackage{dblfloatfix}
\usepackage{hyperref}
\usepackage{tabularx}
\usepackage{pifont}
\usepackage{calc}
\usepackage{caption}
\usepackage{graphicx}
\usepackage{subfig}
\usepackage{float}
\usepackage[ruled,vlined]{algorithm2e}
\usepackage{stackengine}
\usepackage{booktabs}
\usepackage{xcolor}
\usepackage{tabularx}
\usepackage{placeins}
\usepackage{multirow}
\usepackage{makecell}
\usepackage[export]{adjustbox}
\usepackage{tabto}
\usepackage{threeparttable}
\usepackage{printlen}
\usepackage{graphicx,wrapfig,lipsum}
\usepackage{soul}
\usepackage{ragged2e}
\usepackage{wrapfig}
\usepackage[resetlabels,labeled]{multibib}
\usepackage{color, colortbl}
\usepackage[most]{tcolorbox}
\usepackage{booktabs}
\usepackage[table]{xcolor}
\usepackage{float}
\usepackage{stackengine}
\usepackage{xcolor}
\usepackage{tabularx}
\usepackage{placeins}
\usepackage{multirow}
\usepackage{makecell}
\usepackage{hhline}
\usepackage[export]{adjustbox}
\usepackage{tabto}
\usepackage{printlen}
\usepackage{hyperref}
\usepackage{multirow}
\usepackage{makecell}  
\usepackage{amsmath}
\usepackage{listings}

\usepackage{longtable}

\definecolor{algobg}{RGB}{246,244,238}      
\definecolor{rulegray}{RGB}{120,120,120}

\lstdefinestyle{hlsalgo}{%
  language=C++,
  basicstyle=\ttfamily\footnotesize,
  columns=fullflexible,
  keepspaces=true,
  showstringspaces=false,
  backgroundcolor=\color{algobg},
  xleftmargin=0em,
  xrightmargin=0em,
  aboveskip=0.6em,
  belowskip=0.2em,
  frame=tb,
  framexleftmargin=10pt,
  rulecolor=\color{black},
  framerule=0.5pt,
  framesep=6pt,
  lineskip=2pt,
  moredelim=[is][\color{black}]{<O>}{</O>},
  moredelim=[is][\color{pragmaRed}\bfseries]{<R>}{</R>},
  moredelim=[is][\color{pragmaRed}]{<CR>}{</CR>},
  moredelim=[is][\color{pragmaGreen}\bfseries]{<G>}{</G>},
  moredelim=[is][\color{pragmaGreen}]{<CG>}{</CG>},
  moredelim=[is][\color{pragmaMagenta}\bfseries]{<M>}{</M>},
  moredelim=[is][\color{pragmaMagenta}]{<CM>}{</CM>},
  moredelim=[is][\color{pragmaBlue}\bfseries]{<B>}{</B>},
  moredelim=[is][\color{pragmaBlue}]{<CB>}{</CB>},
  moredelim=[is][\color{orange}]{<OR>}{</OR>},
  moredelim=[is][\color{orange}\bfseries]{<COR>}{</COR>},
}

\newcommand{\framework}{Memoryless\xspace}

\definecolor{dmblue}{HTML}{0b03fc}

\definecolor{darkgreen}{RGB}{0,100,0}

\newcommand{\addcites}[1]{\textcolor{red}{~[XXX]}}

\definecolor{lightgray}{gray}{0.95}

\lstdefinestyle{shortlststyle}{
    backgroundcolor=\color{algobg},
    basicstyle=\ttfamily\footnotesize,
    frame=tb,                
    rulecolor=\color{black},
    framerule=0.5pt,
    xleftmargin=0pt,
    xrightmargin=0pt,
    aboveskip=0.5em,
    belowskip=0.5em,
    moredelim=[is][\color{pragmaGreen}\bfseries]{<R>}{</R>},
}
\begin{document}




\setcopyright{acmlicensed}  





\title{Don’t let your Memory defy you: Fragmentation-Aware Serverless Allocation with Elastic Memory Locality}

\settopmatter{authorsperrow=3}

\author{Achilleas Tzenetopoulos}
\affiliation{%
  \institution{National Technical University of Athens}
  \city{Athens}
  \country{Greece}
}

\author{Dimosthenis Masouros}
\affiliation{%
  \institution{National Technical University of Athens}
  \city{Athens}
  \country{Greece}
}

\author{Sotirios Xydis}
\affiliation{%
  \institution{National Technical University of Athens}
  \city{Athens}
  \country{Greece}
}

\author{Francky Catthoor}
\affiliation{%
  \institution{National Technical University of Athens}
  \city{Athens}
  \country{Greece}
}

\author{Dimitrios Soudris}
\affiliation{%
  \institution{National Technical University of Athens}
  \city{Athens}
  \country{Greece}
}



\begin{abstract}


Serverless platforms commonly rely on bundled, memory-centric configurations, where CPU capacity follows the specified memory size.
Resource decoupling reduces this waste, but can create external fragmentation by producing diverse CPU–memory shapes that leave residual capacity stranded across nodes.
Memory disaggregation and tiering can turn these resource holes into usable capacity by enabling elastic memory locality, where instances use different mixes of local and remote memory.
In this paper, we present \textit{\framework{}}, a fragmentation-aware resource manager that exposes memory locality as a serverless control-plane primitive. 
\framework{} jointly selects and places function variants that differ in their compute allocation and memory-locality ratio. 
This allows the scheduler to match instance shapes to fragmented node capacity while preserving SLOs.
We implement \framework{} on top of Knative and evaluate it with trace-driven workloads. 
Compared to state-of-the-art serverless placement frameworks, \framework{} reduces node usage by up to 40\% in steady state and 46\% at peak, while keeping SLO violations within 3\%. 
It also reduces CPU stranding by 44\%, showing that elastic memory locality can convert fragmented node capacity into usable resources.

\end{abstract}

\maketitle


\acmConference[]{}{}{}

\section{Introduction}

Serverless computing platforms such as AWS Lambda~\cite{awslambda} and Microsoft Azure Functions~\cite{azurefunctions} have become widely adopted, due to their ability to abstract away infrastructure management and resource provisioning from end users~\cite{fouladi2019laptop,jonas2017occupy,mampage2022holistic}.
In serverless, developers submit function code, while the provider handles provisioning, placement, scaling, and request dispatching.
This model enables fine-grained pay-per-use execution and allows cloud providers to multiplex many short-lived functions over shared infrastructure~\cite{jonas2019cloud,liu2023demystifying}.

Despite this provider-managed execution model, resource underutilization remains common in serverless clusters~\cite{sahraei2023xfaas, gunasekaran2020fifer}.
A key source for this waste is \textit{internal fragmentation}, where resources are reserved for a function instance but remain unused during execution.
Internal fragmentation mainly arises because serverless platforms expose memory-centric resource configurations, where the selected memory size determines the overall resource shape of a function instance.
For example, in AWS Lambda, users configure only the allocated memory, while vCPUs scale proportionally with the selected memory size~\cite{lambda-bundling,awspricing}.
This coupling can force functions into vCPU–memory shapes that do not match their runtime demand, leaving part of the reserved allocation idle.
To address this problem, prior work has proposed decoupling vCPU and memory allocations, allowing functions to be provisioned closer to their observed demand and improving latency–cost tradeoffs~\cite{bilal2023great,wen2024combofunc}.

However, decoupling vCPU and memory allocations does not eliminate fragmentation.
Once functions are provisioned closer to their actual demand, their resource shapes become more diverse. 
This diversity makes placement harder because residual node capacity can become imbalanced across CPU and memory. 
For example, it may have spare CPU but too little memory, or spare memory but too little CPU to match the next function instance.
The remaining capacity is therefore stranded at the node level, causing \textit{external fragmentation}.
Our experiments on real-world serverless production traces show that idealized allocations can leave up to 78\% of memory and 44\% of CPU capacity externally fragmented~\cite{joosen2024serverless}, shifting resource waste from individual allocations to cluster-level placement and packing.


Existing serverless systems primarily focus on resource configuration selection, queue- and workload-aware request routing, dynamic resource adaptation, or orthogonal execution-layer optimizations such as stateful execution abstractions and memory deduplication~\cite{moghimi2023parrotfish,eismann2021sizeless,zhou2022aquatope,wen2022stepconf,li2023golgi,sinha2025alap,copik2024process,qiu2023user,fuerst2023iluvatar,li2023golgi,romero2021llama}. 
Function placement, however, remains comparatively underexplored, as it is typically delegated to provider-controlled schedulers. 
AWS Lambda~\cite{awslambda}, for example, has been shown to use bin-packing-like placement using bundled CPU--memory to maximize node memory utilization~\cite{wang2018peeking}; this simplifies scheduling, but can waste resources when a function's CPU and memory demands are asymmetric. 
Mu~\cite{mittal2021mu} moves beyond fixed bundles by considering decoupled CPU and memory availability during placement, while ComboFunc~\cite{wen2024combofunc} jointly selects heterogeneous function variants and node placements. 
Nevertheless, these systems still operate over restricted function shapes. 
Mu places fixed per-function configurations, and ComboFunc provides only limited variant flexibility through vCPU quotas, without explicitly accounting for load-dependent latency SLO compliance or goodput.
 

Emerging memory-disaggregation and memory-tiering technologies create an opportunity to turn these resource holes into usable capacity. 
Byte-addressable memory tiers, memory pools, and remote-memory interfaces such as the Compute Express Link (CXL) interconnect standard~\cite{cxl} allow memory capacity to extend beyond node-local DRAM, enabling function instances to use a mix of local and remote memory~\cite{li2023pond,maruf2023tpp,qi2025chrono,berger2025octopus,alverti2025cxlfork,lee2023memtis}. 
This flexibility introduces a new resource-configuration knob by making memory locality \textit{elastic}. 
Serverless platforms can vary not only how much memory an instance receives, but also how much of that memory should consume scarce local DRAM. 
Existing memory-tiering and serverless memory-management systems exploit this capability through page migration, page offloading, remote cloning, or memory reclamation~\cite{maruf2023tpp,lee2023memtis,qi2025chrono,xu2024faasmem,alverti2025cxlfork,nikolos2024squeezy}. 
However, these mechanisms operate primarily at the OS, runtime, or architecture level, below the serverless control plane. 
They manage pages, memory regions, or instance state after an application has already been configured and placed.
As a result, they cannot adapt the memory-locality configuration to dynamic conditions such as fluctuating load or changing resource availability, and the control plane remains agnostic of the local-to-remote memory ratio, unable to select a memory-locality variant that fits a specific resource hole. 

\noindent\textbf{Our work.} 
We present \framework{}, a fragmentation-aware serverless resource manager that exposes memory locality as a control-plane primitive.
Users submit function code and a Service Level Objective (SLO), while the platform reasons over alternative resource configurations for executing that function. 
\framework{} augments this configuration space with an \textit{elastic memory ratio} which determines how much of the function’s memory footprint must consume local DRAM.
This produces multiple latency-feasible variants for the same function, each with a different CPU allocation, memory footprint, and local-memory demand. 
At runtime, Memoryless jointly selects and places these variants to satisfy load while fitting residual node capacity. 
This allows the scheduler to fill resource holes that fixed CPU–memory configurations would leave unused, reducing the active-node footprint while preserving latency constraints.

Overall, this paper makes the following contributions:

\noindent$\blacktriangleright$\textbf{Memory tiering as a control-plane primitive.} 
We identify local and remote memory ratios as a first-class provisioning dimension for serverless orchestration, expanding the configuration space beyond fixed memory–vCPU pairs and enabling new efficiency–performance tradeoffs.

\noindent$\blacktriangleright$\textbf{SLO-aware configuration planning.}
We construct Pareto-optimal function variants over tail latency, throughput, and resource footprint. 
By separating per-instance latency feasibility from aggregate throughput capacity, our planner prunes the design space sequentially, enabling efficient online decisions without sacrificing near-Pareto optimality.


\noindent$\blacktriangleright$\textbf{Placement-aware orchestration of heterogeneous variants.}
We design an orchestration framework that jointly selects, scales, and places heterogeneous function variants to increase function density and reduce resource fragmentation and node footprint under dynamic load.

We implement \framework{} as a prototype on top of Knative and evaluate it using production traces from real-world serverless platforms~\cite{joosen2024serverless}, across diverse functions and function mixes. 
Compared to ComboFunc~\cite{wen2024combofunc}, \framework{} reduces the number of nodes required for deployment by up to 40\% at steady state and 46\% at peak load. 
By packing functions more effectively, \framework{} reduces stranded CPU by up to 44\%, while keeping SLO violations below 3\%.

\section{Background \& Motivation}
\label{sec:motivation}


\subsection{Understanding Fragmentation in Serverless}
Serverless platforms execute large numbers of short-lived function instances over shared clusters~\cite{jonas2019cloud,liu2023demystifying}. 
To improve utilization, providers colocate these instances on the same physical nodes and multiplex CPU and memory across tenants and function instances.
To expose this model to users, providers offer a discrete set of predefined CPU--memory bundles that function owners select from at deployment time.
However, the resource management decisions underlying this model give rise to two forms of resource waste~\cite{li2022serverless}, which we refer to as \textit{internal} and \textit{external fragmentation}.


\subsubsection{Internal Fragmentation}
Internal fragmentation refers to resources reserved for a function instance but left unused during execution.
Figure~\ref{subfig:internal-fragmentation} illustrates this case. 
Functions $f_1$ and $f_2$ reserve larger CPU and memory allocations than their runtime demand, so part of each allocation remains idle.
Although this capacity is physically present on the node, it is enclosed within existing allocations and cannot be used for $f_3$. 
As a result, $f_3$ spills over to a new node.


Despite resource overprovisioning, a main reason for internal fragmentation is resource bundling. 
Commercial serverless platforms expose fixed pairs of vCPUs and memory, limiting users’ ability to provision resources proportional to their actual needs.
In AWS Lambda~\cite{awslambda}, for example, CPU is allocated proportionally to memory, whereas Google's Cloud Run functions~\cite{gcpfunctions} tie vCPU to memory in fixed ratios (e.g., 1 vCPU$\leftrightarrow$2 GB, 2 vCPUs$\leftrightarrow$4 GB, etc.), regardless of whether workloads are compute- or memory-bound.
As a result, increasing memory to satisfy a function’s footprint may also allocate additional vCPU that the function does not need. 
Conversely, increasing vCPU for a compute-bound function may require selecting a larger memory configuration, even if the additional memory remains unused.
Recent studies advocate decoupling CPU and memory provisioning to reduce overprovisioning while preserving performance~\cite{bilal2023great}.
Together, these two causes make internal fragmentation difficult to avoid under fixed bundle models, motivating proposals to decouple CPU and memory allocations.
However, improving per-instance efficiency changes the distribution of allocation shapes, which can increase node-level residual imbalance and hence external fragmentation.

\subsubsection{External Fragmentation}
\begin{figure}[t]
    \centering

    \subfloat[Internal\label{subfig:internal-fragmentation}]{
        \includegraphics[width=0.3\linewidth]
        {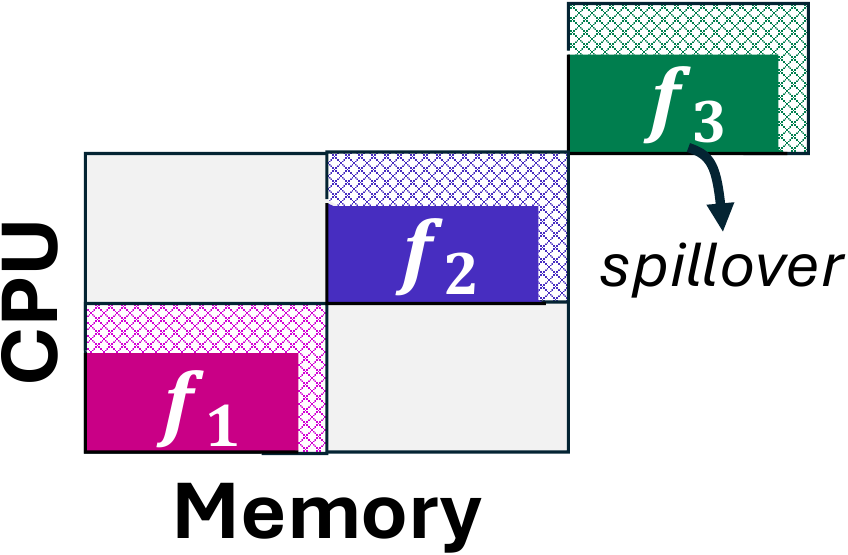}
    }
    \subfloat[External\label{subfig:external-fragmentation}]{
        \includegraphics[width=0.32\linewidth]
        {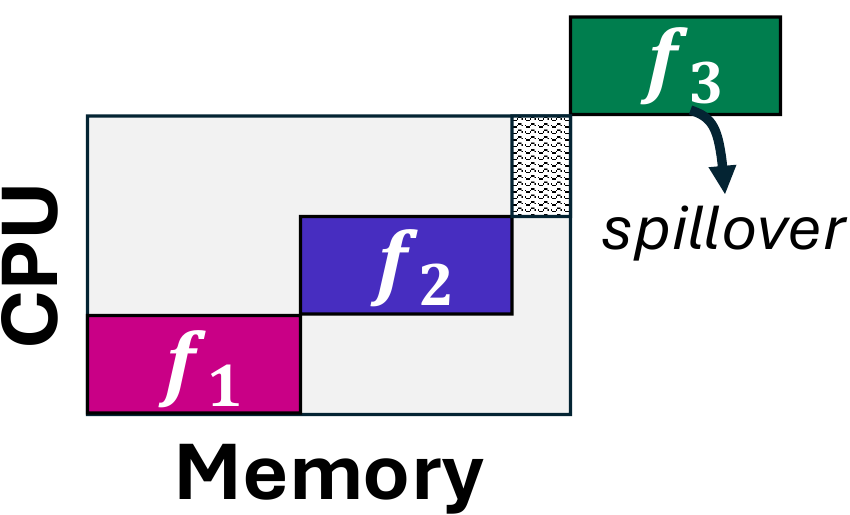}
    }
    \subfloat[Elastic\label{subfig:remote-memory-fragmentation}]{
        \includegraphics[width=0.32\linewidth]
        {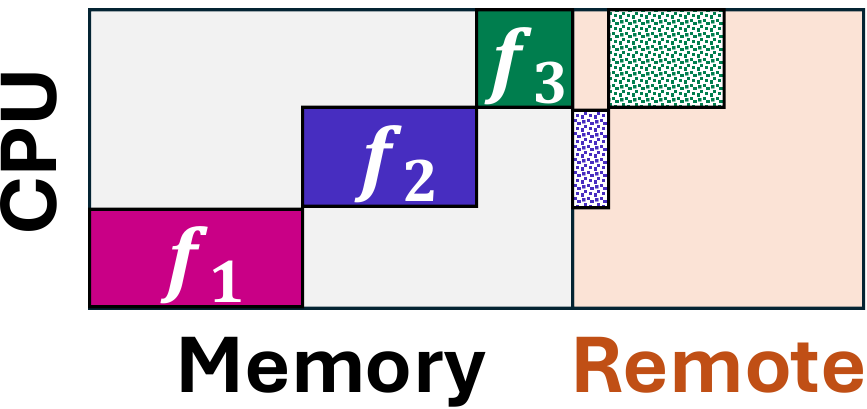}
    }
    \caption{The cases of internal (a) and external (b) fragmentation and elastic memory allocation (c). Decoupling vCPU and memory allocations trades internal fragmentation for external fragmentation. Elastic memory allocation reduces this effect by improving packing efficiency.}
    \label{fig:background-fragmentation}
\end{figure}

To tackle internal fragmentation, prior work proposes decoupling CPU and memory allocations~\cite{wen2024combofunc,bilal2023great}. 
This allows functions to be provisioned closer to their actual resource demands, instead of being selected from fixed CPU-memory bundles.
While this reduces slack inside each function allocation, it still does not entirely eliminate fragmentation.
External fragmentation can still appear at the node level. 
Once CPU and memory are allocated independently, function instances can have more diverse resource shapes.
As shown in Figure~\ref{subfig:external-fragmentation}, the already placed functions leave residual CPU and memory on the node, but the remaining capacity does not match the requirements of $f_3$. 
The node, therefore, has unused resources, yet no feasible placement exists for the next instance. 
As a result, $f_3$ spills over to a new node.
This form of waste is different from internal fragmentation. 
The unused capacity is not trapped inside an oversized function allocation. 
Instead, capacity is stranded at the node level because the residual CPU and memory resources are imbalanced, and no admissible function shape can fit them.



\subsection{Quantifying fragmentation in production traces}

\begin{figure}[t]
	\centering
    \centering
    \subfloat[CPU and memory utilization distributions under bundled memory offerings. Low utilization indicates internal fragmentation.\label{fig:internal}]{
    \includegraphics[width=0.9\columnwidth]{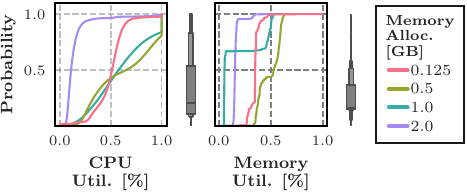}
    }

    \subfloat[External Fragmentation of bundled vs. decoupled resource provisioning.\label{fig:external}]{
\includegraphics[width=0.9\columnwidth]{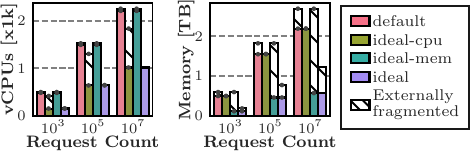}
    }
     \caption{Resource fragmentation in serverless production clusters~\cite{joosen2024serverless}. (a) Using fixed-size bundles leads to overprovisioning and internal fragmentation, while (b) heterogeneous allocations can lead to node-level (external) fragmentation.}
     \label{fig:fragmentation}
\end{figure}

To quantify the effect of internal fragmentation in practice, we analyze the Huawei serverless trace~\cite{joosen2024serverless}.
The trace reports each function’s provisioned resource configuration together with its observed CPU and memory usage over time. 
We estimate internal fragmentation by comparing the resources reserved for each function instance against the resources it actually uses.
Figure~\ref{fig:internal} reports the utilization distributions for the four most frequently used memory configurations\footnote{Configurations are offered as fixed bundles of vCPUs and memory.}.
The CDFs show per-configuration CPU and memory utilization, while the boxplots summarize utilization across all function shapes. 
Low utilization, therefore, corresponds to high internal fragmentation.
The results show that internal fragmentation is widespread rather than limited to a few outlier functions. 
Across all functions, median CPU and memory utilization are only 20\% and 16\%, respectively. 
Memory utilization is especially low, with more than 90\% of allocations using less than 55\% of their provisioned capacity. 
The effect also varies across memory bundles. 
Larger memory configurations exhibit particularly low CPU utilization, indicating that increasing memory often reserves compute capacity that functions do not use. 
At the same time, smaller configurations do not eliminate memory waste, showing that fixed bundles cannot consistently match the resource demand of diverse functions.


We further quantify external fragmentation through a trace-driven placement simulation. 
We derive each function’s CPU and memory demand from the Huawei trace and place function instances on nodes.
We use BestFit as a representative packing-oriented placement heuristic, consistent with prior observations that serverless platforms use placement policies that aim to improve VM utilization~\cite{wang2018peeking}.
New nodes are activated when no existing node can fit the next instance.
We compare four allocation policies: (i) the default bundled configurations in the trace (\emph{default}); (ii) idealized allocations matching actual utilization (\emph{ideal}); and (iii) partially idealized policies that fix either CPU or memory to utilized values (\emph{ideal-cpu}, \emph{ideal-mem}).
Figure~\ref{fig:external} reports the total provisioned CPU and memory as the request volume increases. 
Hatched regions denote external fragmentation, corresponding to residual node resources that remain unused because the remaining CPU and memory capacity on a node are imbalanced. 
The default bundled policy provisions substantially more resources overall, but creates relatively little external fragmentation because its fixed shapes are easier to pack. 
In contrast, idealized decoupling reduces the resources requested by each instance, but produces more diverse CPU--memory shapes that are harder to place. 
As a result, stranded capacity increases substantially, reaching up to 78\% memory fragmentation under ideal-mem and 44\% CPU fragmentation under ideal-cpu. 
These results show that CPU--memory decoupling reduces internal fragmentation, but without placement-aware shape selection, it can shift waste to cluster-level external fragmentation.

\subsection{Elastic Memory Locality as a Resource Knob}


Even with decoupled CPU and memory provisioning, each function instance must reserve sufficient local DRAM for its peak footprint to avoid Out-of-Memory failures.
This rigid locality constraint limits packing flexibility and reintroduces waste as internal slack or node-level external fragmentation.


Emerging memory-disaggregation technologies, such as CXL-backed pooled memory, relax the coupling between memory capacity and physical locality~\cite{li2023pond}. 
They allow a function's footprint to span local DRAM and remote pooled memory, turning locality into a provisioning dimension rather than a fixed constraint. 
While prior work studies disaggregated memory for capacity expansion and cost efficiency~\cite{tzenetopoulos2024disaggregated,wahlgren2023quantitative,masouros2023adrias}, its use as a serverless control-plane lever remains unexplored.
For serverless, locality becomes a spectrum where a function's footprint can be split across heterogeneous memory tiers in tunable proportions.
This reduces local DRAM pressure and improves packing flexibility, while introducing a latency cost that must be managed under SLO constraints.

\begin{figure}
    \centering
    \includegraphics[width=\columnwidth]{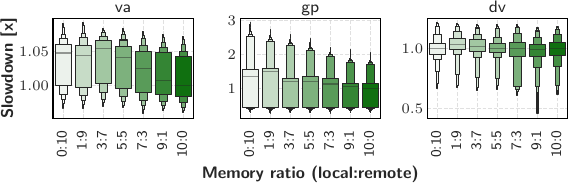}
    \caption{Latency slowdown distribution with varying elastic memory ratio. Slowdown is normalized to the local-only configuration.}\label{fig:motivation-locality}
\end{figure}

This flexibility exposes locality as a tunable provisioning knob alongside vCPU quotas. 
We capture this choice with the \emph{elastic memory ratio (EMR)}, defined as the fraction of an instance's footprint that must be served from local DRAM, while the remainder may be served from the disaggregated pool. 
An EMR of 1.0 corresponds to conventional fully local provisioning, whereas lower values relax local-capacity constraints and create alternative function shapes with smaller local memory footprints. 
Because reducing locality can affect performance, a lower EMR may require a higher vCPU quota to preserve latency or more replicas to sustain throughput~\cite{tzenetopoulos2025towards}. 
Each feasible operating point, therefore, defines a distinct shape in terms of vCPU quota, local memory footprint, and aggregate capacity, expanding the feasible placement region and admitting placements that would otherwise fail under strict locality constraints.
This moves beyond homogeneous function instances~\cite{wen2024combofunc}, where a single variant serves all invocations, and motivates controlled instance heterogeneity as a first-class serverless control-plane mechanism.
Figure~\ref{fig:background-fragmentation} shows that allocating a fraction of $f_2$'s and $f_3$'s memory footprint in remote memory reduces their local DRAM demand and avoids spillover to a new node.

Remote memory accesses introduce higher latency than local DRAM, so locality must be tuned under performance constraints.
To quantify this effect, we deploy three functions, \textit{video-analytics} (va), \textit{graph-pagerank} (gp), and \textit{dna-visualization} (dv), using the setup in \S\ref{subsec:experimental-setup}\footnote{Future CXL Type-3 devices may improve over our PMem-based emulation in latency and bandwidth, making our setup a conservative approximation of CXL-backed memory expansion.}.
For each function, we vary the local-to-remote memory ratio from 0:10, fully remote, to 10:0, fully local, and report invocation-latency slowdown normalized to the local-only configuration.
Figure~\ref{fig:motivation-locality} shows that sensitivity to remote memory is highly function-dependent.
Different functions exhibit varying sensitivity to remote memory allocation due to differences in memory intensity, cache locality, and access patterns~\cite{katsaragakis2025performance}. 
While \textit{video-analytics} and \textit{dna-visualization} remain close to local-only performance across most ratios, with tail-latency differences around 2\%, \textit{graph-pagerank} exhibits substantially higher slowdown, reaching 49\% under remote-heavy configurations due to its more memory-sensitive access pattern.
The memory footprint scales proportionally with the applied ratio; 
thus becoming a configurable knob that directly influences latency.
Thus, reducing local DRAM demand can improve placement flexibility, but EMR must be treated as a performance-sensitive knob, applied selectively to functions with low remote-memory sensitivity or compensated through additional resources when needed.


In practice, serverless configuration is often guided by structured profiling over a user-specified configuration space. 
Tools such as AWS Lambda Power Tuning~\cite{powertune} evaluate functions under user-provided, representative inputs, or an expected workload distribution, and expose cost--latency tradeoffs for configuration selection. 
This workflow assumes that performance can be characterized offline before steady-state deployment.
EMR fits naturally into the same process by extending the profiling space with locality settings, exposing both the latency impact of remote memory and the corresponding function shapes.
As with existing tuning, EMR profiling can be performed offline, with its cost amortized across deployments and repeated invocations.
Overall, EMR turns memory locality into a controllable configuration dimension, enabling placement-friendly function variants while requiring workload-aware profiling to preserve latency and throughput constraints.

\section{Memoryless Overview}
\label{sec:overview}
We propose \textit{\framework{}}, a fragmentation-aware resource manager for serverless environments, that exposes memory locality as an additional control knob.
\framework{} first decouples CPU and memory provisioning, allowing functions to allocate these resources independently.
It further extends this flexibility with elastic memory locality, allowing each function instance to use a mix of local and remote memory.
This expanded configuration space creates function variants with different CPU and local-memory footprints, enabling the control plane to fill fragmented node capacity that would otherwise trigger spillover to additional nodes.

\noindent At a high level, Memoryless relies on three key design points:
\begin{itemize}[leftmargin=*, topsep=0pt]
   \item \textbf{Memory elasticity as a control-plane primitive:}
    \framework{} exposes memory locality as a scheduling decision, allowing function instances to use different local-to-remote memory allocation ratios. 
    This lets the control plane trade locality for placement flexibility.
    \item \textbf{Pareto-optimal configuration generation:}
    \framework{} profiles configurations across CPU allocation, worker parallelism, and memory-locality ratio, retaining only Pareto-optimal variants that capture latency--resource--throughput tradeoffs.
   \item \textbf{Fragmentation-aware allocation and placement of function instances:} 
   Memoryless jointly selects function variants and places instances to reduce residual CPU and memory fragmentation across nodes. 
\end{itemize}

\begin{figure*}[t]
    \centering
    \includegraphics[width=\textwidth]{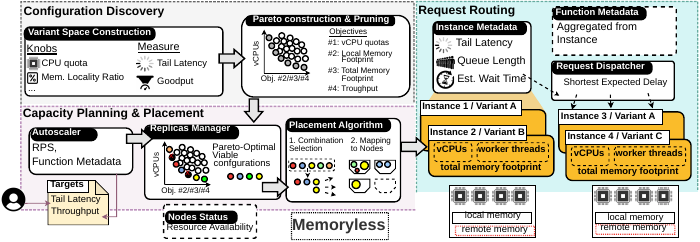}
    \caption{Memoryless Overview.}\label{fig:memoryless-overview}
\end{figure*}

\framework{} operates entirely within the provider control plane, selecting resource allocations, memory-locality configurations, and placements to meet tenant SLOs while improving provider-side resource efficiency.
Tenants provide function source code and, optionally, workload information such as representative input classes or expected load regimes.
Similar to existing profiling tools~\cite{powertune}, \framework{} profiles each function across configuration knobs such as vCPU limits, worker parallelism, and elastic memory ratio (EMR).
Tenants therefore express performance and cost intent through SLOs, such as tail-latency targets, rather than manually specifying explicit \mbox{(vCPU, memory)} configurations~\cite{zhou2022aquatope}.

Each profiled variant is assigned a performance envelope capturing tail latency and sustainable throughput, i.e., the maximum request rate it can serve within the SLO violation threshold.
\framework{} then prunes variants dominated in latency, throughput, or resource footprint, yielding a Pareto-optimal set that bounds runtime search and enables controlled instance heterogeneity.

At runtime, \framework{} scales function instances to meet latency constraints under incoming load.
It jointly selects and places variants from the latency-pruned Pareto set whose aggregate capacity can serve the observed request rate, guided by each node's residual CPU and memory availability.
This enables tighter fits, improving packing density and reducing resource fragmentation.

\section{Memoryless Design}
\label{section:design}

Figure~\ref{fig:memoryless-overview} illustrates the architecture of \framework{}.
The control plane consists of three main phases, namely \textit{Configuration Discovery}, \textit{Capacity Planning and Placement}, and \textit{Request Routing}.
\textit{Configuration Discovery} runs offline and produces the Pareto-optimal variant set used at runtime.
\textit{Capacity Planning and Placement} reacts to workload changes by selecting instance counts, variants, and placements.
Finally, the \textit{Request Dispatcher} routes requests across the deployed heterogeneous instances using runtime metadata, such as queue length and latency estimates.

\subsection{Configuration Discovery}
\label{subsec:configuration-discovery}

\begin{figure}[!h]
    \centering
    \includegraphics[width=\columnwidth]{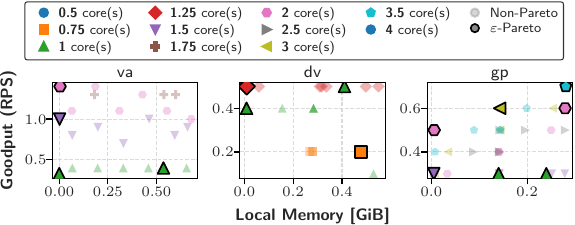}
    \caption{Pareto-optimal per-function configurations under varying EMR and vCPU quotas. The frontier minimizes cores and local memory footprint while maximizing goodput (RPS).}\label{fig:motivation-pareto}
\end{figure}

\framework{} models each function as a set of execution \textit{variants} spanning vCPU allocation, worker count, and elastic memory ratio (EMR). 
These knobs jointly determine tail latency, sustained throughput, and two dimensioned resource footprint, yielding a combinatorial configuration space. Following autotuning workflows such as AWS Lambda Power Tuning~\cite{powertune}, users provide representative payloads and their relative frequencies; \framework{} profiles variants on tail-latency, throughput, and resource usage (CPU and local/remote memory footprint).
Our multi-objective optimization depends on (i) sustained goodput maximization and (ii) resource footprint minimization along three dimensions: vCPU allocation\footnote{this indirectly translates to memory bandwidth}, local DRAM footprint, and total-memory footprint\footnote{Tail latency is measured during profiling but enforced later via SLO-based pruning.}. 
We include total-memory footprint because increasing worker parallelism can raise an instance's overall memory demand.
We apply $\epsilon$-Pareto pruning to suppress near-equivalent variants arising from profiling noise or negligible footprint differences.
Figure \ref{fig:motivation-pareto} plots goodput (vertical axis) against local memory footprint (horizontal axis) across different vCPU quotas, with worker count fixed at one,\footnote{Configurations with multiple workers are included in the exploration space but omitted here for clarity.} for the profiled variants of three functions. 
Highlighted points denote Pareto-optimal configurations that maximize goodput for a given memory footprint. 
Across functions, non-dominated points expose different resource shapes from distinct vCPU--EMR combinations, including iso-goodput configurations, such as in \textit{graph-pagerank}, with different resource footprints.

\begin{figure}
    \centering
    \subfloat[Pareto construction using qEHVI.]{%
        \includegraphics[width=0.49\columnwidth]{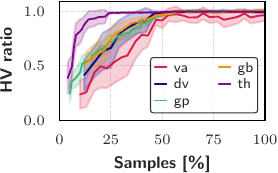}
        \label{fig:qehvi}%
    }\hfill
    \subfloat[Pareto optimal configurations vary with input size.]{%
        \includegraphics[width=0.47\columnwidth]{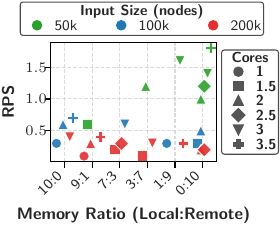}%
        \label{fig:input-pareto}%
    }
    \caption{Pareto-optimal function variant-set construction with (a) efficient sampling, and (b) input awareness.}
    \label{fig:profiling}
\end{figure}

To explore this space efficiently, \framework{} uses hypervolume improvement-based Bayesian optimization with qEHVI~\cite{daulton2020differentiable}, applied greedily over a discrete candidate pool.
The acquisition function targets the Pareto frontier by favoring configurations expected to increase dominated hypervolume, improving goodput without increasing resource footprint across the optimized dimensions.
Since the search space is a discrete cross-product of cores, workers, and EMR, \framework{} scores unmeasured candidates from this pool and selects the most promising configurations for profiling.
As shown in Figure~\ref{fig:qehvi}, the hypervolume ratio saturates near 1.0 with limited samples, confirming that most of the Pareto tradeoff surface is recovered early. 
When input characteristics vary significantly, the set of dominant configurations may shift accordingly.
Figure \ref{fig:input-pareto} illustrates the Pareto-optimal configurations for different payload sizes for \textit{graph-pagerank}. 
To capture this variation, \framework{} constructs separate Pareto frontiers per input bucket, retaining only non-dominated variants across all buckets to produce the final candidate set.

Although tail latency and throughput are jointly determined under load, their feasibility conditions have fundamentally different structure that \framework{} exploits for efficient pruning. 
Tail-latency SLO compliance is an instance-level property, independent of deployment scale, allowing non-compliant variants to be eliminated beforehand.
Throughput feasibility, in contrast, is aggregate, depending on the combined goodput of deployed instances.
In this regime, queuing delay is kept small so that latency is dominated by service time; the control plane scales the deployment to maintain low queuing and allow aggregate capacity to grow with the number of instances.
\framework{} exploits this asymmetry to prune sequentially. 
Pareto filtering followed by latency pruning yields a compact set of SLO-feasible variants that bounds the runtime search space. 
At runtime, the \textit{Capacity Planner} reasons exclusively over this set to provision sufficient aggregate goodput and select placements that minimize fragmentation, matching resource shapes to residual node capacity without sacrificing goodput, while supporting timely decisions over heterogeneous instance mixes.


\subsection{Capacity Planning \& Placement}
\label{subsec:capacity_planning}
This phase covers runtime decisions regarding scaling, resource allocation, and placement.

\subsubsection{Autoscaler}
\label{subsunsec:autoscaler}

The autoscaler tracks load fluctuations and provides the control plane with an aggregate capacity target that preserves latency headroom.
We target reaction times on the order of a few seconds. 
At this granularity, serverless workloads can be highly bursty, and current forecasting techniques are often insufficiently accurate over such short horizons~\cite{joosen2023does,dehigama2024composing}; we therefore adopt a reactive scaling strategy; predictive scaling is orthogonal and can be layered on top.  
Unlike utilization-based autoscalers, \framework{} requires an explicit goodput target coupled with queuing-delay feedback to support joint variant selection and placement. 

At each step, it computes:
\begin{equation}
\small
C_{\text{target}} = \frac{\max(A, D)}{\rho_{\text{eff}}} + f(B),
\label{eq:autoscaler}
\end{equation}
where $A$ and $D$ are smoothed arrival and completion rates, $B$ is the backlog (queued requests), $\rho_{\text{eff}}$ is an adaptive utilization target, and $f(B)$ is a backlog-driven correction term.
The first term estimates the steady-state capacity required to track the offered load; $\max(A,D)$ provides robustness to short-term measurement noise and transient imbalance between arrivals and completions, while dividing by $\rho_{\text{eff}}$ maintains headroom and avoids operating at full saturation.
$\rho_{\text{eff}}$ is bounded within a fixed range $[\rho_{\min}, \rho_{\max}]$ and is adjusted based on observed tail queuing delay. 
As latency approaches the SLO, $\rho_{\text{eff}}$ is reduced to create additional headroom.
The second term $f(B)$ is a monotonically increasing correction that temporarily raises $C_{\text{target}}$ above the steady-state estimate when backlog accumulates, provisioning enough capacity to drain queued requests within a bounded time horizon.
For stability, the autoscaler smooths inputs, enforces cooldown periods, caps per-step changes, and disables downscaling while backlog remains non-zero.
The resulting $C_{\text{target}}$ is passed to the joint variant selection and placement component.

\subsubsection{Joint Variant Selection \& Placement}
\label{subsubsec:joint_placement}




\begin{algorithm}[t]
\caption{Memoryless Variant Selection and Placement}
\label{alg:small-holes-first}
\small

\KwIn{
Nodes $\mathcal{N}$ with residual $(\mathrm{cpu}^{\mathrm{free}}_n,\mathrm{mem}^{\mathrm{free}}_n)$\;
Pareto-optimal SLO-feasible variants $\mathcal{P}$ with $(c_v,m_v^{\mathrm{loc}},G(v))$\;
Incremental capacity requirement $\Delta\lambda>0$\;
}
\KwOut{Placements $\Pi$ and achieved added goodput $\hat{\lambda}$.}

$\Pi \leftarrow \emptyset;\ \hat{\lambda} \leftarrow 0$\;

$N \leftarrow \mathrm{sort}(\mathcal{N},\mathrm{hole}(n))$\tcp*[l]{smallest residual capacity first}

$V \leftarrow \mathrm{stable\_sort}(\mathcal{P},s(v))$\tcp*[l]{best footprint-per-goodput first}

\ForEach{$n \in N$}{
  \While{$\hat{\lambda} < \Delta\lambda-\epsilon$}{
    $v \leftarrow \mathrm{first}\{v\in V:\ c_v \le \mathrm{cpu}^{\mathrm{free}}_n \land m_v^{\mathrm{loc}}\le \mathrm{mem}^{\mathrm{free}}_n\}$\;

    \If{$v=\bot$}{\textbf{break}\tcp*[l]{no feasible variant fits on $n$}}

    $\Pi \leftarrow \Pi \cup \{(n,v)\}$\;

    $(\mathrm{cpu}^{\mathrm{free}}_n,\mathrm{mem}^{\mathrm{free}}_n) \mathrel{-}= (c_v,m_v^{\mathrm{loc}})$\;

    $\hat{\lambda} \leftarrow \hat{\lambda} + G(v)$\;
  }

  \If{$\hat{\lambda} \ge \Delta\lambda-\epsilon$}{\textbf{break}\tcp*[l]{capacity target reached}}
}
\Return $\Pi,\hat{\lambda}$\;
\end{algorithm}

\framework{} performs planning and placement incrementally to limit churn and its overheads, such as cold starts~\cite{kondrashov2025high}, and disruption to in-flight requests.
Given the incremental capacity signal $\Delta\lambda$ from the autoscaler, the planner jointly selects heterogeneous variants and places resulting instances onto nodes, considering only the Pareto-optimal SLO-feasible variant set (\S\ref{subsec:configuration-discovery}).
When input buckets are used (see AWS Power Tuning~\cite{powertune} config file), $\Delta\lambda$ is apportioned according to the workload mix.
\footnote{Input buckets are used only when input characteristics materially shift dominance relations; otherwise a single frontier suffices.}
Each candidate instance must fit on a node subject to vCPU and \textit{local} DRAM constraints; remote memory is provisioned from a shared disaggregated pool and is therefore not constrained by per-node DRAM capacity~\cite{amaro2020can}.
During scale-out, the planner favors placements that pack onto already-active nodes and reduce residual fragmentation; during scale-in, it removes instances to preserve packing quality and increase the likelihood of fully draining nodes.

\textbf{Scale-out:}
Joint variant selection and placement is combinatorial; \framework{} uses a greedy heuristic. 
When ($\Delta\lambda > 0$), nodes are processed in increasing order of hole score (Eq.~\ref{eq:hole-score}), prioritizing nodes with the tightest residual capacity:
\begin{equation}
\small
\mathrm{hole}(n) = \frac{\mathrm{cpu}^{\mathrm{free}}_n}{C^{\mathrm{node}}} + \frac{\mathrm{mem}^{\mathrm{free}}_n}{M^{\mathrm{node}}},
\label{eq:hole-score}
\end{equation}
where $(\mathrm{cpu}^{\mathrm{free}}_n, \mathrm{mem}^{\mathrm{free}}_n)$ are the remaining vCPU and local-memory resources on node $n$, and $(C^{\mathrm{node}}, M^{\mathrm{node}})$ are per-node capacities.
For each node, the algorithm (Alg.~\ref{alg:small-holes-first}) repeatedly selects the feasible variant with the best normalized resource usage per unit of goodput and places one instance, updating residual resources after each step:
\begin{equation}
\small
s(v) = \frac{
    \sqrt{\left(\frac{c_v}{C^{\mathrm{node}}}\right)^2 + \left(\frac{m_v^{\mathrm{loc}}}{M^{\mathrm{node}}}\right)^2}
}{
    \max(G(v), \epsilon)
},
\label{eq:variant-score}
\end{equation}
where $c_v$, $m_v^{\mathrm{loc}}$, and $G(v)$ are the vCPU demand, local-memory demand, and sustained goodput of variant $v$, and $\epsilon$ is a small constant for numerical stability.
The algorithm runs in $O(N\log N + P\log P + (N+K)P)$, where $N$ is the number of nodes, $P$ the number of Pareto variants, and $K$ the number of instances added.

\textbf{Scale-in:}
When $\Delta\lambda < 0$, \framework{} performs a symmetric greedy scale-in procedure that incrementally removes instances.
Nodes are processed in decreasing order of hole score (i.e., most underutilized nodes first) encouraging the system to consolidate load and drain nodes entirely.
Within each node, instances are removed in increasing order of sustained goodput, enabling fine-grained capacity reduction.
The process continues until the target reduction $|\Delta\lambda|$ is achieved or no further safe removals remain, limiting disruption and unnecessary churn.

\subsection{Request Dispatcher}
\label{subsec:dispatcher}


\framework{} employs an asynchronous request dispatcher that routes requests across the active set of heterogeneous function instances.
Since variants differ in resource configuration (vCPU allocation, EMR, worker parallelism) and therefore in service capacity and latency characteristics, each variant is exposed as a separate service endpoint.
The dispatcher maintains per-function routing state that is updated whenever instances are added or removed, ensuring routing decisions reflect the current deployment.


Requests are routed using a Shortest Expected Delay (SED) policy that accounts for instance heterogeneity.
For each arriving request, the dispatcher identifies its input class $c$ and prioritizes variants deployed for that class, consistent with the class-specific capacity planning in \S\ref{subsec:capacity_planning}.
Each instance is modeled as a multi-server queue with a fixed number of parallel workers, tracked via a min-heap of next-free timestamps.
The expected delay for variant $v$ is estimated as the sum of the predicted queueing delay, derived from the earliest available worker, and the expected service time from the variants' latency profile.
The request is routed to the variant with minimum predicted delay, naturally balancing load according to each instance's capacity and performance.
Requests that exceed a timeout proportional to the SLO target, are dropped.


Cold (warming) instances are never assigned requests directly.
When all warmed instances exceed a queue threshold while new instances are still warming, dispatch is temporarily delayed and retried after a short backoff, preventing overload while allowing new instances to become eligible.
The dispatcher continuously collects runtime signals, including queueing delay, end-to-end latency, queue length, and per-variant throughput.
Variants that repeatedly violate latency targets are temporarily blacklisted from routing and planning, while sustained wait time and backlog growth are forwarded to the autoscaler for scaling decisions.
\section{Evaluation}
\label{sec:evaluation}

\subsection{Evaluation Methodology}
We evaluate \framework{} using trace-driven serverless workloads deployed on Kubernetes and Knative.
Our evaluation focuses on the operating-system-visible abstraction of remote memory as a slower, byte-addressable memory tier, rather than on any specific interconnect implementation (e.g., CXL).
This abstraction lets us study how elastic memory allocation affects placement, fragmentation, and node usage independently of the details of a particular remote-memory fabric.
Overall, our evaluation aims to answer the following research questions:

\noindent\textbf{RQ1:} How effectively does \framework{} reduce node usage under trace-driven workloads?

\noindent\textbf{RQ2:} Do multi-function deployments amplify the benefits of \framework{} compared to single-function deployments?

\noindent\textbf{RQ3:} How does \framework{} achieve node reduction in practice, in terms of resource fragmentation and node packing density?


\noindent\textbf{RQ4:} How close does the online heuristic come to offline optimal selection and placement?

\subsubsection{Experimental Setup}
\label{subsec:experimental-setup}
We deploy 16 virtual machines\footnote{One of which serves as the control plane and runs on a separate server.} on servers equipped with an Intel\textsuperscript{\textregistered} Xeon\textsuperscript{\textregistered} Gold 5218R CPU @ 2.10\,GHz.
Each VM is exposed to Kubernetes as a separate worker node.
Unless otherwise stated, each worker node advertises 4 allocatable vCPUs and 2\,GB of allocatable local DRAM to Kubernetes.
Within each VM, local memory is exposed as a NUMA node with vCPUs backed by DRAM.
In addition, we configure five memory-only NUMA nodes backed by Intel Optane Persistent Memory, which we use to emulate remote memory with higher access latency than local DRAM.
This setup lets us vary the local-to-remote memory ratio from fully local to predominantly remote memory while retaining standard NUMA-based placement and allocation control.
Kubernetes is used for container orchestration, with Knative providing the serverless execution environment.
Intel Optane Persistent Memory is configured in Device Direct Access (\texttt{devdax}) mode with \texttt{memory-ram} semantics, making the PMem-backed NUMA nodes transparently usable by unmodified applications.
All experiments run on Linux kernel 6.9.
We use the latest version of \texttt{numactl}\footnote{\url{https://github.com/numactl/numactl}} that supports weighted NUMA interleaving, enabling fine-grained control over page allocation across DRAM and PMem-backed nodes.

\subsubsection{Benchmarks}
\label{subsubsec:benchmarks}

We use five representative serverless functions drawn from widely used benchmark suites~\cite{vswarm,copik2021sebs}. Specifically, we consider:
\textit{Thumbnailer (th)}, which resizes input images;
\textit{PageRank (gp)}, which computes webpage rankings using the PageRank algorithm;
\textit{Graph BFS (gb)}, which generates a graph and executes the Breadth-First Search (BFS) algorithm;
\textit{Video Analytics (va)}, which runs an object detection model (SqueezeNet~\cite{iandola2016squeezenet}) on video input; and
\textit{DNA Visualization (dv)}, which transforms DNA sequences into interactive visualizations.

\begin{figure*}[h]
    \centering
    \includegraphics[width=0.9\textwidth]{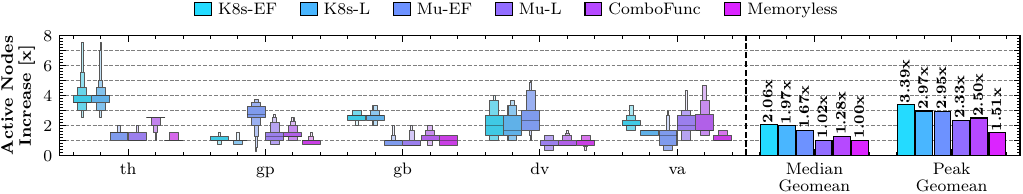}
    \caption{Active-node distribution and peak node usage for different function deployments and scheduling strategies using the Huawei Serverless invocation trace~\cite{joosen2024serverless}. Normalized per function on the minimum median value.}\label{fig:per-function-nodes}
\end{figure*}

\begin{figure*}[t]
	\centering
    \centering
    \subfloat[Active Nodes count distribution.\label{fig:mix_active_nodes}]{
\includegraphics[width=0.4\textwidth]{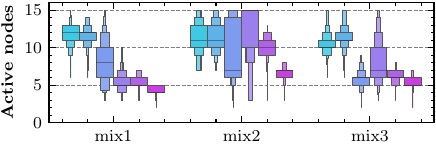}
    }\hfill
        \subfloat[Active Nodes Summary.\label{fig:mix_active_nodes_summary}]{
\includegraphics[width=0.25\textwidth]{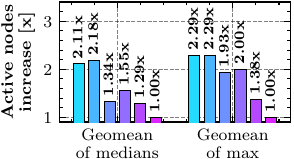}
    }\hfill
    \subfloat[SLO violations.\label{fig:slo_violations}]{
\includegraphics[width=0.165\textwidth]{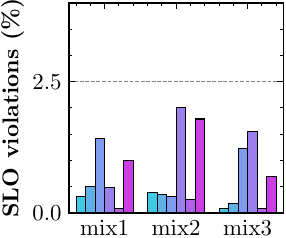}
    }
    \caption{Active-node distribution and SLO violations per strategy for function mixes deployments using the Huawei Serverless invocation trace~\cite{joosen2024serverless}.}
     \label{fig:comparative_analysis_mixes}
\end{figure*}

\subsubsection{Workload Traces}
\label{sec:eval:traces}
We use the workload traces provided by Huawei~\cite{joosen2024serverless}. 
Since the trace exposes only opaque function identifiers and the exact workloads are unknown, we relabel the most invoked subset of functions with the benchmark functions used in our evaluation. 
For each function, we crop the trace to a fixed duration of 10 minutes. 
To match the capacity of our cluster, we scale down the offered load (originally collected from hyperscale deployments) by capping per-function request rates via sliding-window–based probabilistic thinning~\cite{law2007simulation}, which preserves original timestamps and temporal structure.
For the workload mixes, we randomly select per-function RPS caps such that the aggregate upper-bound RPS fits within our testbed capacity.
For a given experiment, all baselines replay the same thinned trace, with identical timestamps, request inputs, and per-function caps.

\subsubsection{Baselines}
We compare \framework{} with:

\textbf{Kubernetes (K8s) -EF/-L:}
We use the default Kubernetes scheduler (used by Knative) as the state-of-practice baseline.
It uses the \textit{LeastAllocatedResources} policy for node selection.
In our implementations, K8s uses homogeneous function variants selecting either the ones with the minimum latency (K8s-L) or the ones with maximum resource efficiency (K8s-EF), minimizing resource usage per goodput.


\textbf{Mu~\cite{mittal2021mu} (-EF/-L):}
We implement a Mu-style placement baseline.
At each placement epoch, after the autoscaler determines the set of pods to schedule, Mu orders pods by their dominant resource share~\cite{ghodsi2011dominant} and places them greedily using the heuristics.
We evaluate two homogeneous-variant settings, selecting the variant that minimizes resource usage per unit goodput (EF) or minimizes latency (L).

\textbf{ComboFunc~\cite{wen2024combofunc}:}
We implement the placement algorithm of ComboFunc.
Specifically, we use heterogeneous function instances selected from a Pareto-optimal set, constructed by varying the vCPU count and the worker count.
While no further information was provided in the paper, we have used the Autoscaling, Capacity Planning, and Request Routing approach of \framework{}.



These baselines isolate the impact of variant heterogeneity, capacity planning, placement strategy, and remote-memory variants.

\subsubsection{Oracle Formulations}
In addition to deployable baselines, we compare against ILP-based oracle formulations that compute per-function optimal placements. 
These ILPs are solved offline on small problem instances using CBC~\cite{cbc} via PuLP~\cite{pulp}, and serve as reference points for solution quality; they are not suitable for online control-plane operation.

\textbf{ILP-N (Min-Nodes):}
Selects variant combinations per node to meet the target throughput, while minimizing the total number of active nodes. 
Stage 1 minimizes active nodes; Stage 2 (with node count fixed) minimizes throughput overshoot and total instance count.
The formulation assumes stateless replacement of the functions' deployment.

\textbf{ILP-NC (Min-Nodes + Min-Churn):}
Extends ILP–N by modeling the current state of deployment.
For each node/variant, decides how many instances to keep and add.
The two-stage structure is preserved. Stage 1 minimizes active nodes; Stage 2 minimizes deployment churn (additions + removals), followed by throughput overshoot and instance count.
This formulation captures transition cost while preserving the optimal node footprint.

\subsection{Comparative Analysis}

\textbf{RQ1. Single-function deployment.}
We first deploy each function independently using Huawei invocation traces~\cite{joosen2024serverless} (\S\ref{sec:eval:traces}) and compare \framework{} against each baseline.
Figure~\ref{fig:per-function-nodes} shows the distribution of active node count over time for each function, where a node is active if it hosts at least one function instance.
Strategy rankings vary across functions due to differences in remote-memory sensitivity, candidate-variant diversity, and the selected allocation shape for homo- geneous-instance baselines.
The right panel reports the geometric mean across functions of median and maximum active-node count.
Median active nodes capture typical node demand when nodes can be dynamically lent, borrowed, or reclaimed, whereas the maximum captures peak footprint under coarser-grained node management.
Across these single-function deployments, \framework{} reduces median active-node usage by 28\% compared to ComboFunc~\cite{wen2024combofunc}. 
Although Mu-L~\cite{mittal2021mu} achieves competitive median-case node usage, it requires 54\% more nodes at peak, indicating weaker robustness to transient load spikes.

\begin{figure*}[t]
    \centering
    \includegraphics[width=0.9\textwidth]{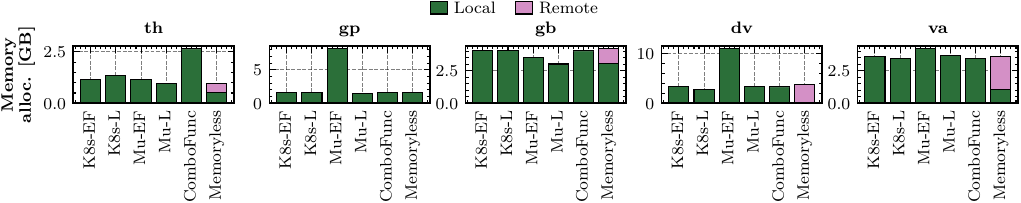}
    \caption{Median local and remote memory allocation across functions and placement strategies.}\label{fig:per-function-remote-memory}
\end{figure*}

\textbf{RQ2. Function-mix deployment.}
To evaluate behavior under increased shape diversity, we deploy three multi-function mixes drawn from the five functions, with per-function RPS caps sampled randomly subject to the aggregate upper-bound constraint.
All strategies replay the same traces and are evaluated by active-node count and SLO violation rate, defined as the fraction of requests that are dropped or whose end-to-end latency, including queuing, exceeds the target.
Figure~\ref{fig:mix_active_nodes} shows active-node usage over time.
\framework{} consistently outperforms all baselines, achieving up to 40\% lower median and 46\% lower peak node usage compared to ComboFunc, and up to 50\% and 67\% lower median and peak node usage compared to the best homogeneous strategy, Mu-EF.
The K8s load-balancing baseline consistently spreads load across many nodes, resulting in high active-node usage across all mixes.
Regarding SLO adherence, ComboFunc achieves the lowest violation rate, below 1\%, but at the cost of higher node usage, while \framework{} maintains a worst-case violation rate of 2.9\% in mix2.
These larger gains under function mixes show that workload diversity creates opportunities for shape-aware consolidation, as \framework{} can combine variants with complementary CPU, memory, and locality profiles on fewer nodes.

\textbf{RQ3. Mechanism behind node reduction.}
We next analyze how selective remote-memory use, variant diversity, and tighter residual-capacity packing explain \framework{}'s node reductions,
As shown in Figure~\ref{fig:per-function-remote-memory}, \framework{} uses remote memory selectively rather than uniformly across functions, indicating that it does not rely on remote memory indiscriminately.
For example, in \textit{dna-visualization (dv)}, 98.5\% of memory is allocated remotely, while in \textit{graph-pagerank (gp)}, all pages are allocated locally, consistent with the high remote-memory sensitivity of gp shown in Figure~\ref{fig:motivation-locality}.
This behavior indicates that \framework{} exploits remote memory only when the additional placement flexibility outweighs the locality overhead.

\begin{figure}[t]
    \centering
    \includegraphics[width=0.85\columnwidth]{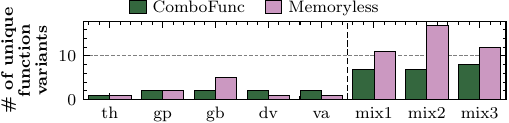}
    \caption{Number of unique function variants used for the single-function and mix deployments.}\label{fig:unique-variants}
\end{figure}

\framework{}'s node reduction also comes from broader variant diversity under function mixes. 
In single-function workloads, \framework{} uses 1.83 variants on average, close to ComboFunc~\cite{wen2024combofunc}, the only other baseline with heterogeneous variants. 
Under function mixes, however, this gap widens substantially: as shown in Figure~\ref{fig:unique-variants}, \framework{} uses up to 16 variants across functions, covering a larger portion of the Pareto frontier. 
This broader variant usage lets \framework{} better match the heterogeneous latency--throughput--resource tradeoffs of mixed workloads.


\begin{figure*}[t]
    \centering
    \includegraphics[width=0.9\textwidth]{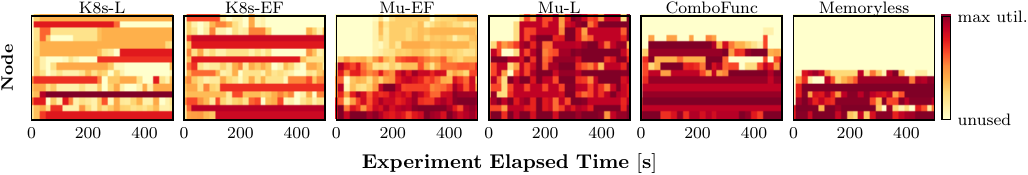}
    \caption{Node occupancy over time across placement strategies. Rows represent nodes, columns represent elapsed time, and color encodes each node's allocated-to-capacity ratio. Memoryless consolidates load onto fewer, fuller nodes, reducing partially utilized capacity relative to the baselines.}\label{fig:node-fullness}
\end{figure*}

This broader variant diversity translates directly into tighter packing. 
Figure~\ref{fig:node-fullness} shows node occupancy over time for mix2.
Rows correspond to nodes, columns to elapsed time, and color encodes each node's allocated-to-capacity ratio, computed as the root mean square (RMS) of normalized CPU and memory allocations.
K8s baselines immediately spread instances across all nodes, resulting in low per-node density.
Mu-L, which relies on ``\textit{larger}'' instances, allocates more resources per node but leaves more residual capacity unused.
ComboFunc and \framework{} pack instances more efficiently, with \framework{} further increasing node density by relaxing local-memory constraints through remote-memory variants.

Quantitatively, \framework{} achieves 23.7\% higher packing efficiency than ComboFunc and 105--125\% higher efficiency than K8s.
Figure~\ref{fig:fragmentation} confirms that these denser placements reduce residual waste.
Across the three function mixes, \framework{} achieves the lowest fragmentation in both local memory and CPU, reducing median local-memory fragmentation by 6--58\% and CPU fragmentation by 44--92\%.
Notably, \framework{} allocates 44\% more CPU resources than ComboFunc, yet reduces median active-node usage by 29\%, showing that the additional CPU allocation is absorbed within a tighter overall placement.

\begin{figure}[t]
    \centering
    \includegraphics[width=0.9\columnwidth]{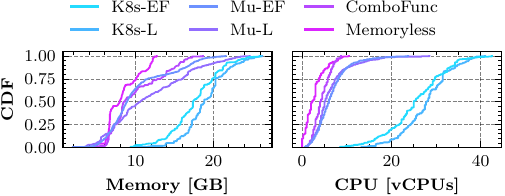}
    \caption{CDF of fragmented (stranded) node resources over time for the function mix deployments.}\label{fig:fragmentation}
\end{figure}

    




\subsection{Goodput Stress Test}

Beyond minimizing active nodes for a target workload trace~\cite{joosen2024serverless}, we include a complementary goodput stress test that measures how much load a fixed-size cluster can sustain before saturation.
All strategies use the same node budget, SLO target, and autoscaling policy.
We progressively increase the arrival rate of the \textit{thumbnailer} function until saturation, defined as the point where the combined SLO violation and request drop rate exceeds 5\%.
We define goodput as the maximum sustainable arrival rate below this threshold.

\begin{figure}[t]
    \centering
    \includegraphics[width=0.9\columnwidth]{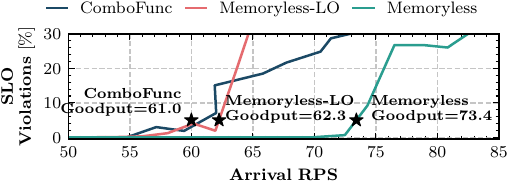}
    \caption{Maximum goodput achieved by each strategy for the \textit{thumbnailer} function, defined as the highest RPS before SLO violations exceed 5\%.}\label{fig:goodput}
\end{figure}

Figure~\ref{fig:goodput} compares ComboFunc, Memoryless-LO, and \framework{}.
Memoryless-LO is a local-only version of \framework{}, so both ComboFunc and Memoryless-LO are restricted to local-memory variants, whereas \framework{} can also select remote-memory variants.
Memoryless-LO improves goodput by approximately 4\% over ComboFunc, showing that \framework{}'s placement and variant-selection strategy provides benefits even without remote memory.
Enabling elastic memory locality further increases goodput by an additional 18\% over Memoryless-LO.
These results indicate that memory-locality elasticity can translate the improved resource utilization observed earlier (Figure~\ref{fig:fragmentation}) into higher sustainable throughput under fixed cluster capacity. 



\subsection{Scalability  \& Optimality Gap Analysis}
\label{subsec:evaluation-optimality}

To answer RQ4, we quantify how close \framework{} comes to offline-optimal selection and placement while remaining usable online.
We compare \framework{} in simulation against two ILP formulations that are computationally prohibitive for online deployment, and against \textit{ComboFunc+}, a ComboFunc variant that also supports remote-memory variants.
The experiment consists of 300 sequential placement decisions under low- and high-intensity Poisson arrivals, selecting and placing instances of \textit{th}, \textit{gb}, and \textit{gp}.

\begin{figure}[h]
    \centering


    \subfloat[CDF of Nodes Used\label{subfig:ecdf_nodes}]{
        \includegraphics[width=0.95\columnwidth]{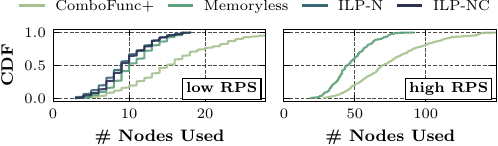}
    }
    
    
    \subfloat[CDF of Decision Latency\label{subfig:ecdf_time}]{
        \includegraphics[width=0.95\columnwidth]{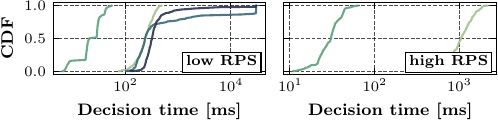}
    }
    
    \subfloat[CDF of Churn\label{subfig:churn}]{
        \includegraphics[width=0.96\columnwidth]{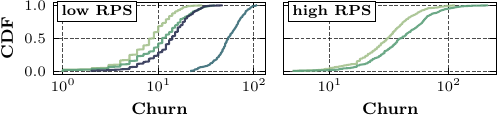}
    }

    \caption{Trade-offs among node efficiency, decision latency, and churn across RPS intensities. \framework{} closely tracks ILP oracles while remaining viable for online control.}
    \label{fig:optimal_gap}
\end{figure}

Figure~\ref{fig:optimal_gap} shows the results.
Despite operating online, \framework{} closely tracks the ILP oracles, remaining within 10--15\% of \textit{ILP-N} in average node count and within 6\% in peak node count (Fig.~\ref{subfig:ecdf_nodes}), whereas \textit{ComboFunc+} over-allocates nodes by 33--40\%.
\textit{ILP-NC} reduces function-instance churn, measured as instance additions and removals, compared to \textit{ILP-N} (Fig.~\ref{subfig:churn}), because \textit{ILP-N} aggressively rearranges existing instances at each step to improve packing.
However, both ILP formulations exceed the 3-minute solver timeout under high RPS, making them unsuitable for online deployment.


\framework{} trades moderately higher churn than \textit{ComboFunc+} ($\sim$20--40\% on average) for better consolidation, using $\sim$35\% fewer nodes.
It also achieves online-compatible decision latency, with p95 decision times in the tens of milliseconds rather than the seconds-to-minutes required by ILP approaches (Fig.~\ref{subfig:ecdf_time}).
Unlike \textit{ComboFunc+}, whose placement procedure reaches second-level decision latency under high RPS, \framework{} maintains stable decision times as demand increases.
Overall, \framework{} approaches offline-optimal node efficiency while maintaining practical churn and online-compatible decision latency.



\section{Related Work}

Prior work on serverless spans provisioning, orchestration, SLO-aware execution, and runtime-level memory management. 
We focus on the control-plane orchestration systems most closely related to our work and discuss memory tiering and resource adaptation as complementary directions.

\subsection{Serverless provisioning and control-plane orchestration}

A large body of prior work focuses on selecting resource configurations to minimize cost or meet latency SLOs for functions~\cite{moghimi2023parrotfish,eismann2021sizeless}, or at the granularity of serverless workflows using profiling or learning-based techniques~\cite{zhou2022aquatope,mahgoub2022orion,wen2022stepconf,zhang2024jolteon,bhasi2021kraken}. 
While effective at meeting per-stage SLOs, these systems operate as user-level mechanisms atop managed platforms and are therefore constrained by provider-defined resource bundles, which limit their ability to jointly apply fragmentation-aware allocation and placement, exploiting elasticity through function instance and memory heterogeneity.

Other works introduce heterogeneity through per-invocation adaptation~\cite{sinha2025alap,tzenetopoulos2026omegakypous,li2023golgi,kondrashov2025melding}. 
ALAP~\cite{sinha2025alap} delays configuration selection until invocation inputs are available; $\Omega$kypous~\cite{tzenetopoulos2026omegakypous} dynamically scales core and uncore frequencies per workflow stage based on available slack and input size; 
Golgi~\cite{li2023golgi} uses instance heterogeneity for queue-aware request routing. 
These approaches primarily focus on invocation-level right-sizing, leaving auto-scaling, allocation and placement decisions to the underlying platform scheduler.

A smaller set of systems explicitly considers placement-aware orchestration~\cite{liu2024harmonizing,wen2024combofunc,mittal2021mu, fuerst2022locality, abad2018package}. 
In~\cite{abad2018package}, authors schedule functions based on cached package availability, while~\cite{fuerst2022locality} balances invocation locality and server load to reduce cold-start overheads.
Jiagu~\cite{liu2024harmonizing} incorporates placement decisions to improve performance, but assumes fixed per-instance configurations without exposing CPU--memory decoupling as a control-plane dimension. Mu~\cite{mittal2021mu} integrates autoscaling, placement, and request routing to reduce fragmentation under SLOs, but relies on homogeneous instances and local memory allocation. 
ComboFunc~\cite{wen2024combofunc} jointly selects heterogeneous configurations and placements during scaling, showing that configuration diversity can improve responsiveness and utilization. 
However, it optimizes a single cost-oriented objective that does not explicitly capture multi-dimensional fragmentation or reason about SLO adherence under varying load, so cost-optimal choices can still yield inefficient packing, as shown in \S\ref{sec:evaluation}. 
In contrast, our work models allocation as a two-dimensional vCPU--memory problem, orchestrates multiple SLO-feasible operating points per function, and enforces sustained goodput as a constraint, directly targeting fragmentation reduction and improved cluster-level multiplexing.

\subsection{Memory Tiering and Runtime-Level Mechanisms}

A substantial body of work improves utilization by relaxing the coupling between compute and memory, through cluster-level disaggregation~\cite{li2023pond,masouros2023adrias} and process-level tiering across heterogeneous memory tiers~\cite{maruf2023tpp,lee2023memtis}. Within the serverless context, FaaSMem~\cite{xu2024faasmem} adapts these ideas by managing memory placement across initialization and idle phases and migrating pages based on observed access patterns. 
TrEnv~\cite{huang2026trenv} uses CXL- and RDMA-backed memory templates to share initialized function state across nodes and enable efficient restoration of repurposable execution environments.
CXLfork~\cite{alverti2025cxlfork} enables fast process cloning over shared CXL memory and dynamically tiers checkpointed state between local and remote memory.
These systems optimize memory placement, backing, or sharing within function execution environments, but do not jointly treat memory locality as a resource-configuration and placement decision.
A related line of work improves utilization through resource harvesting, dynamically reclaiming or hot-(un)plugging CPU and memory at the VM or container level~\cite{zhang2021faster,fuerst2022memory}, or vertically scaling microVMs~\cite{zhang2024faascale}. 
These mechanisms primarily mitigate transient underutilization within individual instances.
In contrast, our work targets the provisioning and orchestration layer, treating memory capacity, locality, and configuration diversity as explicit control-plane decisions to improve cluster-wide packing.
Both runtime-level memory management and harvesting mechanisms are therefore orthogonal and complementary to our approach.

\section{Conclusion}

We presented \framework{}, a fragmentation-aware resource allocation and placement framework for serverless functions. 
\framework{} expands the Pareto-optimal configuration space with Elastic Memory Ratio and jointly selects, combines, and places heterogeneous instances with flexible resource shapes to improve density and reduce cluster-wide resource waste. 
Implemented on Knative and evaluated on a real hardware cluster, \framework{} reduces active node count by up to 40\% and improves goodput by 22\% over the closest state-of-the-art baseline.













\bibliographystyle{ACM-Reference-Format}
\bibliography{refs}

\end{document}